\documentclass[aps,pre,twocolumn,showpacs,groupedaddress]{revtex4}

\usepackage{graphicx}
\usepackage{dcolumn}
\usepackage{bm}
\usepackage{amssymb}

\begin{document}


\title{Dynamic Modes of Red Blood Cells in Oscillatory Shear Flow}

\author{Hiroshi Noguchi}
\email[]{noguchi@issp.u-tokyo.ac.jp}
\affiliation{
Institute for Solid State Physics, University of Tokyo,
 Kashiwa, Chiba 277-8581, Japan}

\date{\today}

\begin{abstract}
The dynamics of red blood cells (RBCs) in oscillatory shear flow was studied using
differential equations  of  three variables: a shape parameter, 
the inclination angle $\theta$, and phase angle $\phi$ of the membrane rotation.
In steady shear flow, three types of dynamics occur depending on the shear rate and viscosity ratio.
i) tank-treading (TT): $\phi$ rotates while the shape and $\theta$ oscillate.
ii) tumbling (TB): $\theta$ rotates while the shape and $\phi$ oscillate.
iii) intermediate motion: both $\phi$ and $\theta$ rotate synchronously or intermittently. 
In oscillatory shear flow, RBCs show various dynamics based on these three motions.
For a low shear frequency with zero mean shear rate, a limit-cycle oscillation occurs, 
based on the TT or TB rotation at a high or low shear amplitude, respectively. 
This TT-based oscillation well explains recent experiments.
In the middle shear amplitude, RBCs show an intermittent or synchronized oscillation.
As shear frequency increases, 
the vesicle oscillation becomes delayed with respect to the shear oscillation.
At a high frequency, multiple limit-cycle oscillations coexist.
The thermal fluctuations can induce transitions between two orbits at very low shear amplitudes.
For a high mean shear rate with small shear oscillation, 
the shape and $\theta$ oscillate in the TT motion
but only one attractor exists even at high shear frequencies.
The measurement of these oscillatory modes is a promising tool 
for quantifying the viscoelasticity of RBCs, synthetic capsules, and lipid vesicles.
\end{abstract}

\maketitle

\section{Introduction}

Soft deformable objects, such as liquid droplets, vesicles, cells, and synthetic capsules
exhibit various behaviors under flows. 
Among these objects, red blood cells (RBCs) have received a great deal of attention,
since they are important  for both fundamental research and medical 
applications.  
The rheological property of RBCs is one of the main
factors for the flow resistance of blood, since 
the volume fraction of RBCs in normal human 
blood is around $45$\%~\cite{fung04,chie87}.
In patients with diseases such as diabetes mellitus and sickle cell anemia, the deformability of RBCs is
reduced, and RBCs often block the microvascular flow~\cite{fung04,chie87,tsuk01,shel03,higg07}.

In a steady shear flow with flow velocity ${\bf v}=\dot\gamma y {\bf e}_x$, 
fluid vesicles exhibit
i) a tank-treading (TT) mode with a constant inclination angle $\theta$ 
at low viscosity of the internal fluid $\eta_{\rm {in}}$
or low membrane viscosity $\eta_{\rm {mb}}$, ii)
 a tumbling (TB) mode appears at high $\eta_{\rm {in}}$ or 
$\eta_{\rm {mb}}$ with low shear rate $\dot\gamma$,
and iii) a swinging (SW) motion (also called trembling or vacillating-breathing)
at middle $\eta_{\rm {in}}$ or $\eta_{\rm {mb}}$ with high $\dot\gamma$
~\cite{kell82,seif99,pozr03,beau04,nogu04,nogu05,nogu09,made06,misb06,nogu07b,kant06,dank07,lebe07,lebe08,desc09,desc09a,mess09}.
In all of the above three phases, a membrane (TT) rotation occurs expect in the limit 
$\eta_{\rm {in}} \to \infty$ or $\eta_{\rm {mb}}\to \infty$.
The TT-TB transition at low $\dot\gamma$
is described well by the theory of Keller and Skalak 
(KS)~\cite{kell82}, which assumes a fixed ellipsoidal vesicle shape.
At high $\dot\gamma$, the shape deformation is not negligible
and induces shape transitions \cite{nogu04,nogu05,nogu09} and the SW phase \cite{misb06,nogu07b,kant06,dank07,lebe07,lebe08,desc09,desc09a,mess09}.

RBCs \cite{gold72,abka07} and synthetic capsules \cite{chan93,walt01,navo98,rama98,kess08,sui08,bagc09,tsub10} 
also transit from TB to TT with increasing $\dot\gamma$,
and the TT mode is accompanied by (swinging) oscillation of their lengths and $\theta$.
Recently, this behavior was explained by the extended KS theory, where
 the membrane shear elasticity is taken into account as
an energy barrier for the membrane rotation of a phase angle $\phi$~\cite{skot07}.
The angles $\theta$ and $\phi$ are depicted in Fig.~\ref{fig:cart}(a).
More recently, we extended this theory~\cite{skot07} to include the shape deformation of RBCs~\cite{nogu09b}.
In TT, the RBC shape and $\theta$ oscillate with the TT rotation frequency.
Most of the phase behaviors are not qualitatively different between fixed-shape and deformable RBCs.
Synchronized phases of the $\theta$ and $\phi$ rotations with integer ratios of the rotation frequencies
as well as intermittent rotations in the middle ranges of shear rate $\dot\gamma$ 
for both fixed-shape and deformable RBCs were found to exist~\cite{nogu09b}.
For microcapsules with low bending rigidity, which have no saddle point in the free-energy potential,
these coexistence regions of $\theta$ and $\phi$ rotations vanish~\cite{nogu10b}.
Our results show good agreement with recent experiments \cite{abka07} and simulations \cite{kess08,sui08,bagc09,tsub10}.

It is very important to understand the dynamic response of RBCs in time-dependent flows,
since blood flows {\it in vivo} are not steady.
However, the dynamics of RBCs and vesicles in time-dependent flows have been explored far less than in steady flows.
Recently for fluid vesicles,
membrane wrinkling was found after inversion 
of an elongational flow~\cite{kant07,turi08},
and shape or orientational oscillation was observed in structured channels
\cite{nogu10a}.
For RBCs, a shape oscillation in an oscillatory shear flow with $\dot\gamma(t)= \dot\gamma_0 \sin(2\pi f_{\gamma} t)$ was
observed experimentally~\cite{wata06}.
However, the mechanism and fundamental properties of this oscillation are not understood.
Watanabe {\it et al.} investigated the oscillation only in a narrow range of the shear amplitude $\dot\gamma_0$ and frequency $f_{\gamma}$.
We want to address the following questions: does the angle $\theta$ or $\phi$ rotate in the experimental condition? 
How does the oscillation depend on $\dot\gamma_0$ and $f_{\gamma}$?
Can intermittency and synchronization of $\theta$ and $\phi$ rotations exist in oscillatory flow?
Do RBCs approach a single orbit independent of initial states?
In this paper, we applied our phenomenological theoretical model to oscillatory shear flow and
 found that the oscillation in Ref.~\cite{wata06} is a TT-based oscillation, and
several other dynamic modes appear depending on the shear amplitude and frequency.
Understanding these frequency dependence is a basic step to reveal the RBC dynamics in more complicated time-dependent flows such as blood flows {\it in vivo}.
The amplitude of shape oscillations is a useful quantity for evaluating RBC deformability \cite{wata06,wata07a}.
Very recently, we studied the dynamics of fluid vesicles in oscillatory flows using a similar theoretical model with  two variables \cite{nogu10}. The effects of RBC shear elasticity can be understood by the comparison with the results of fluid vesicles.

Under physiological conditions, 
an RBC has a constant volume $V = 94\mu{\rm m}^3$, surface area $S= 135\mu{\rm m}^2$,
 $\eta_{\rm {in}}=0.01$Pa$\cdot$s, $\eta_{\rm {mb}}\sim 10^{-7}-10^{-6}$Ns/m,
 membrane shear elasticity $\mu=6\times 10^{-6}$N/m, and 
bending rigidity $\kappa=2 \times 10^{-19}$J \cite{nogu09,fung04,tran84,dao06,vazi08}.
Hereafter, the model and results are presented with dimensionless quantities (denoted by a superscript $*$).
The lengths and energies are normalized by $R_0=\sqrt{S/4\pi}=3.3$ $\mu$m
and $\varepsilon_0=\mu R_0^2=6.5\times 10^{-17}$J, respectively.
The relative viscosities are
$\eta_{\rm {in}}^*=\eta_{\rm {in}}/\eta_0$
and $\eta_{\rm {mb}}^*=\eta_{\rm {mb}}/\eta_0R_0$, where 
$\eta_0$ is the viscosity of the outside fluid. 
The reduced volume of RBCs is $V^*=V/(4\pi R_0^3/3)=0.64$.
In this paper, 
a typical viscosity of the surrounding fluid in the experiments, $\eta_0= 0.02$Pa$\cdot$s is chosen:
$\eta_{\rm {in}}^*=0.5$ and $\eta_{\rm {mb}}^*=1.55$.
There are three or four intrinsic time units for zero and finite mean shear rate, respectively:
the shape relaxation time $\tau=\eta_0 R_0/\mu$ by the shear elasticity $\mu$;
and the times of shear flows $1/\dot\gamma_0$, $1/\dot\gamma_{\rm m}$, and $1/f_{\gamma}$.
The reduced shear amplitude $\dot\gamma_0^*=\dot\gamma_0 \tau$, 
mean shear rate $\dot\gamma_{\rm m}^*=\dot\gamma_{\rm m} \tau$, 
and shear frequency $f_{\gamma}^*=f_{\gamma}/\dot\gamma_0$ are used.
In typical experimental conditions, the Reynolds number is low, Re$<1$; hence,
the effects of the inertia are negligible.

The generalized KS model for RBCs and the dynamics in steady shear flow are 
briefly described in Sec. \ref{sec:theory} and in Sec. \ref{sec:std}, respectively.
The dynamics for zero and finite mean shear rate are presented in Sec. \ref{sec:osc} and in Sec. \ref{sec:bias}, respectively. The effects of thermal fluctuations at very low shear rates are described in Sec. \ref{sec:noise}.
The summary and discussion are given in Sec. \ref{sec:sum}.

\section{Theoretical Model}\label{sec:theory}

In our phenomenological theoretical model~\cite{nogu09b},
the shape parameter $\alpha_{13}=(L_1-L_3)/(L_1+L_3)$ is employed to describe the shape deformation of RBCs,
where $L_1>L_2$ and $L_3$ are the principal lengths of the RBC on the vorticity ($xy$) plane
and  in the vorticity ($z$) direction, respectively. 
Here, it is assumed that one of the principal axes is in the $z$ direction and
the symmetric axis of RBCs with the thermal-equilibrium discoidal shape is on the $xy$ plane.
In the absence of flow, RBCs have a biconcave discoidal shape with $\alpha_{13}=0$.
The dynamics of a model RBC is described by three differential equations for $\alpha_{13}$, the inclination angle
$\theta$, and phase angle $\phi$,
\begin{eqnarray}
\label{eq:al}
\frac{d \alpha_{13}}{\dot\gamma dt} &=& \Big\{1-\big(\frac{\alpha_{13}}
                   {\alpha_{13}^{\rm {max}}}  \big)^2\Big\}
    \Big\{ -\frac{A_0}{\dot\gamma^*} 
       \frac{\partial F^*}{\partial \alpha_{13}} 
                + A_1\sin(2\theta)\Big\}, \\
\label{eq:qks}
\frac{\ \ d \theta}{\dot\gamma dt} &=& \frac{1}{2}\big\{-1+f_0 f_1 \cos(2\theta)\big\} - \frac{f_0 d \phi}{\dot\gamma dt},\\
\label{eq:phiks}
\frac{\ \ d \phi}{\dot\gamma dt} &=& -\frac{(c_0/\dot\gamma^*V^*) \partial F^*/\partial \phi + \cos(2\theta) }
      {2f_1\{1+f_2(\eta_{\rm {in}}^* -1) 
                  + f_2f_3 \eta_{\rm {mb}}^*\}},
\end{eqnarray}
where  $A_0= 45/2\pi(32+23\eta_{\rm {in}}^*+16\eta_{\rm {mb}}^*)V^*$ and 
$A_1= 60/(32+23\eta_{\rm {in}}^*+16\eta_{\rm {mb}}^*)$.
Factors $f_0$, $f_1$, $f_2$, $f_3$, and $c_0$ are
the functions of the length ratios ($L_2/L_1$, $L_3/L_1$).
A detailed  description of this model is given in Ref.~\cite{nogu09b}.
Equation~(\ref{eq:al}) is derived on the basis of the perturbation theory~\cite{seif99,misb06,lebe07,lebe08} of 
quasi-spherical vesicles~\cite{nogu07b,nogu09b}.
The first and second terms in the last parentheses in Eq.~(\ref{eq:al}) or Eq.~(\ref{eq:phiks}) represent
the RBC elastic forces to recover the thermal-equilibrium state and the external shear stress, respectively.
Equations~(\ref{eq:qks}) and (\ref{eq:phiks}) are given by the extended KS theory in Ref.~\cite{skot07}.
The first and second terms in Eq.~(\ref{eq:qks}) are given by Jeffery's theory \cite{jeff22} for the dynamics of solid objects.
The third term represents the effects of the membrane $\phi$ rotation to the dynamics of the angle $\theta$.

The free energy $F(\alpha_{13}, \phi)$ of RBCs is estimated by
the simulation of the RBC elongation by mechanical forces:
$F^*(\alpha_{13}, \phi)=F^*_1(\alpha_{13}) + F^*_2(\alpha_{13})\sin^2(\phi)$
  with $F^*_1(\alpha_{13})= 5\alpha_{13}^2+ (40/3)\alpha_{13}^3+
  (230/4)\alpha_{13}^4$ and $F^*_2(\alpha_{13})= 0.2+0.8\alpha_{13}$.
The RBC membrane is modeled as a triangular network with
 a bond potential $U_{\rm {bond}}=(k_1/2)(r-r_0)^2\{1+ (k_2/2)(r/r_0-1)^2\}$
and bending potential $U_{\rm {bend}}= (\kappa/2) \int (C_1+C_2)^2 dS$,
where $C_1$ and $C_2$ are the principal curvatures at each point of the membrane.
Our simulation with $\mu=(\sqrt{3}/4)k_1=6\times 10^{-6}$N/m,  $\kappa=2 \times 10^{-19}$J, and $k_2=1$
 reproduces the force-length curves of the optical-tweezers experiment
and previous simulations \cite{dao06,vazi08} very well \cite{nogu09}.

When the free energy $F$ is independent of $\phi$,
the equations describe the dynamics of fluid vesicles.
Here, we only investigate the dynamics of RBCs but the model itself can be applied to other elastic capsules by the modification of $F$.
Note that the equation of $\theta$ in the perturbation theory should not be applied to the dynamics of vesicles at $V^* \lesssim 0.8$, including RBCs ($V^* \simeq 0.6$) \cite{nogu10}.
It gives too low critical viscosity $\eta_{\rm {in}}^*$ of TT-TB transition
(TB motion occurs for fluid vesicles even at $\eta_{\rm {in}}^*=1$ and $\eta_{\rm {mb}}^*=0$).

To investigate the effects of thermal fluctuations in Sec. \ref{sec:noise},
Gaussian white noises $g_{\alpha}(t)$, $g_{\theta}(t)$, and $g_{\phi}(t)$ are added to Eqs. (\ref{eq:al})--(\ref{eq:phiks}), respectively,
 where 
$\langle g_i(t)\rangle = 0$ and
$\langle g_i(t)g_j(t')\rangle = 2 D_i \delta_{i,j}\delta(t-t')$ with $i,j \in  \{\alpha, \theta, \phi \}$.
The fluctuation-dissipation theorem gives
the diffusion coefficients $D_i=k_{\rm B}T/\zeta_i$,
where $\zeta_i$ are friction coefficients and $k_{\rm B}T$ is the thermal energy:
$\zeta_{\alpha}= \varepsilon_0\tau/A_0\{1-(\alpha_{13}/\alpha_{13}^{\rm {max}})^2\}$,
$\zeta_{\theta}= \zeta_{\phi}/f_0$, and $\zeta_{\phi}=2f_1\{1+f_2(\eta_{\rm {in}}^* -1) + f_2f_3 \eta_{\rm {mb}}^*\}V^*\varepsilon_0\tau/c_0$.
Equations without thermal noises are numerically integrated using the fourth-order Runge-Kutta method 
with a time step $\Delta t \leq 0.0005/\dot\gamma_0$ or $\Delta t \leq 0.0005/\dot\gamma_{\rm m}$
for oscillatory flow with zero (Sec. \ref{sec:osc}) or finite (Sec. \ref{sec:bias}) mean shear rate, respectively.
Equations with thermal noises are numerically integrated
using the  second-order Runge-Kutta method with a time step $\Delta t = 0.0002/\dot\gamma_0$ (Sec. \ref{sec:noise}).

\section{Steady Flow} \label{sec:std}

First, we briefly describe RBC dynamics in steady shear flow.
Detailed dynamics is described in Ref.~\cite{nogu09b}.
At a low shear rate $\dot\gamma^*<\dot\gamma^*_{\rm {tb}}$,
RBCs show TB motion, where $\theta$ rotates while $\phi$ oscillates,
since the energy barrier locks the phase angle at $\phi\simeq 0$.
At a high shear rate $\dot\gamma^*>\dot\gamma^*_{\rm {tt}}$,
the TT motion  occurs, where $\phi$ rotates while $\theta$ oscillates.
As $\dot\gamma^*$ increases from $\dot\gamma^*=\dot\gamma^*_{\rm {tb}}$ to $\dot\gamma^*_{\rm {tt}}$,
the rotation frequency ratio increases from $f_{\rm {rot}}^{\phi}/f_{\rm {rot}}^{\theta}=0$ to $1$.
The coupling of $\theta$ and $\phi$ rotations induces synchronization 
with integer ratios of $f_{\rm {rot}}^{\phi}$ and $f_{\rm {rot}}^{\theta}$.
Here, an angle change of $\pi$ is counted as one rotation.
Unsynchronized (intermittent) rotations are obtained between the regions of synchronizations.
This type of synchronization is called the Devil's staircase \cite{berg84}.
In this model, the RBC approaches one attractor from any initial configuration.

As $\eta_{\rm {in}}^*$ increases, both $\dot\gamma^*_{\rm {tb}}$ and $\dot\gamma^*_{\rm {tt}}$
increases. 
At $(\eta_{\rm {in}}^*, \eta_{\rm {mb}}^*)=(0.5, 1.55)$, the critical shear rates are
$(\dot\gamma^*_{\rm {tb}}, \dot\gamma^*_{\rm {tt}})=(0.01615, 0.01831)$.
The TT phase disappears at $\eta_{\rm {in}}^* \gtrsim 0.9$ and  $\eta_{\rm {mb}}^*/\eta_{\rm {in}}^*=3.1$.

The RBC free-energy potential $F$ has a saddle point at $\phi=\pi/2$: 
energy minimum for constant $\phi=\pi/2$
and energy maximum in the $\phi$ rotation for constant $\alpha_{ 13}$.  
This saddle point is observed for RBCs by experiments \cite{fisc04} and simulations \cite{nogu09b}.
It plays a significant role to the phase behavior of RBCs and microcapsules.
When the saddle point vanishes, the coexistence phases of $\theta$ and $\phi$ rotations disappear \cite{nogu10b}.
Kessler {\it et al.} contested that intermittent rotation is an artifact of the theoretical model~\cite{kess08},
since they did not observe it in their simulations.
However, it would be caused by the low bending rigidity of their quasi-spherical capsule model.

\begin{figure}
\includegraphics{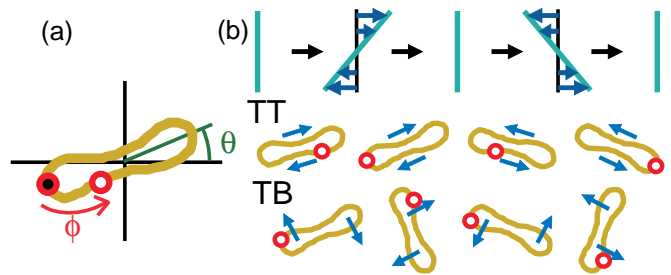}
\caption{ \label{fig:cart}
(Color online)
Schematic of a red blood cell (RBC) in oscillatory shear flow.
(a) Inclination angle $\theta$ and phase angle $\phi$.
(b) Tank-treading (TT) based oscillation ($\phi$ rotates back and forth),
and tumbling (TB) based oscillation ($\theta$ rotates back and forth).
}
\end{figure}

\begin{figure}
\includegraphics{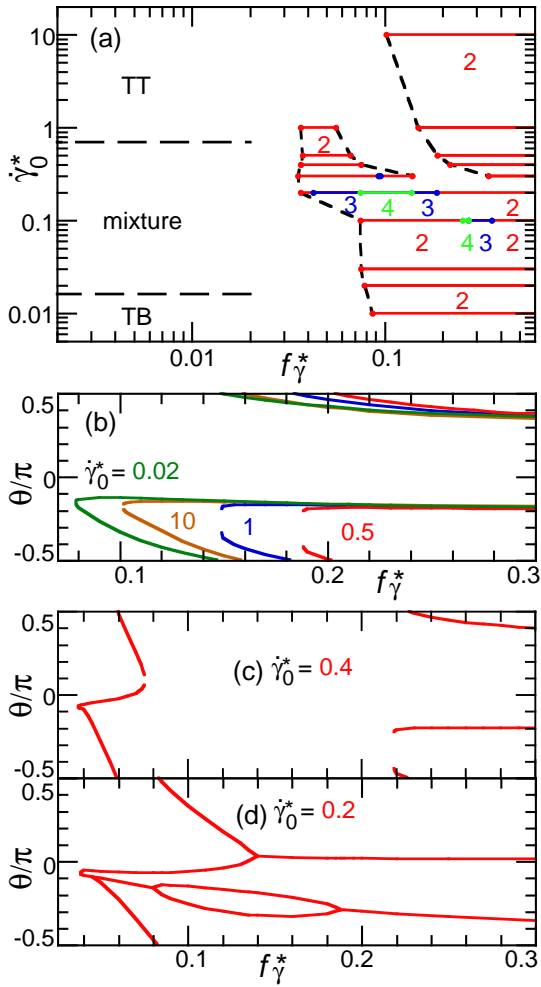}
\caption{ \label{fig:atline}
(Color online)
RBC dynamics in oscillatory shear flow with zero mean shear rate.
(a) Dynamic phase diagram.
(b)--(d) Domain boundary of limit-cycle oscillations at various $\dot\gamma_0^*$.
Each domain consists of 
the initial positions $(\alpha_{13}, \theta, \phi)=(0, \theta_{\rm i}, 0)$ at $t=0$
approaching the same attractor.
For low shear frequency $f_{\gamma}^*$,
TT- or TB- based oscillation occurs at low or high shear amplitude $\dot\gamma_0^*$, respectively.
In the middle regions, intermittent or synchronized oscillations appear.
For high $f_{\gamma}^*$, multiple attractors exist.
Solid lines in (a) represent two (red), three (blue), and four (green) attractors
obtained from the domains in (b)--(d). 
Dashed lines are visual guides.
}
\end{figure}

\begin{figure}
\includegraphics{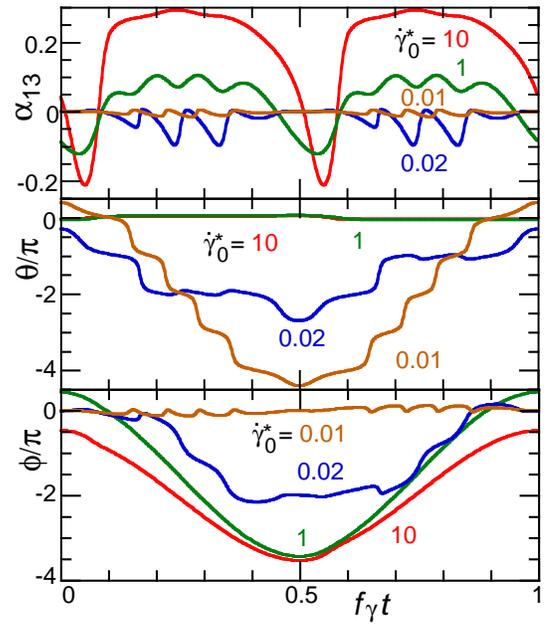}
\caption{ \label{fig:rq5}
(Color online)
Limit-cycle oscillations for various 
shear amplitude $\dot\gamma_0^*$
at low shear frequency $f_{\gamma}^*=0.005$.
Only one limit cycle exists for each $\dot\gamma_0^*$.
}
\end{figure}

\begin{figure}
\includegraphics{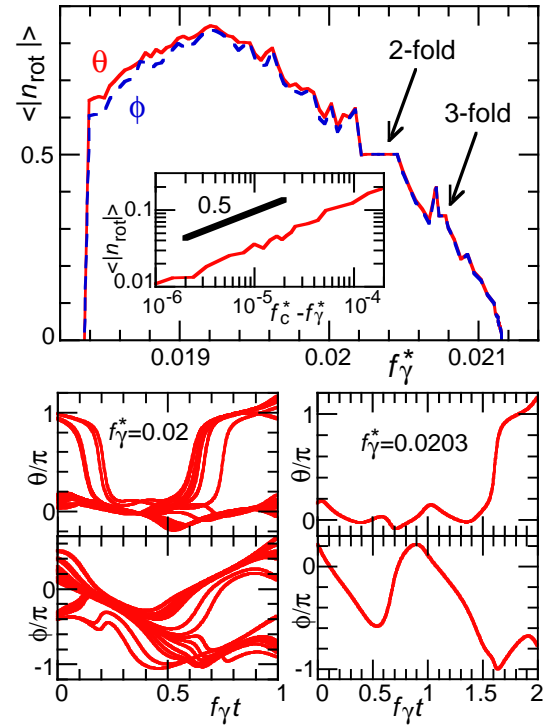}
\caption{ \label{fig:itc}
(Color online)
RBC dynamics for the low shear frequencies $f_{\gamma}^*$ and middle shear amplitude $\gamma_0^*=0.02$.
Top panel: average number $\langle n_{\rm {rot}} \rangle$ of rotations per shear-oscillation period $1/f_{\gamma}$.
Bottom panels: time evolution of $\theta$ and $\phi$ in intermittent or $2$-fold limit-cycle oscillation
at $f_{\gamma}^*=0.02$ or $0.0203$, respectively.
The inset of the top panel shows
the log-log plot with the critical frequency $f_{\rm c}^*=0.02111584$.
The error bars are smaller than the line thickness.
}
\end{figure}

\section{Oscillatory Flow with Zero Mean Shear Rate}\label{sec:osc}

In the oscillatory shear flow with $\dot\gamma(t)= \dot\gamma_0 \sin(2\pi f_{\gamma} t)$,
much more complicated dynamics occurs depending on $\dot\gamma_0^*$ and $f_{\gamma}^*$ than in the steady flow.
The phase diagram is shown in Fig.~\ref{fig:atline}.
The RBC approaches either one or multiple attractors in the limit $t\to \infty$ 
depending on the initial positions in the phase space ($\alpha_{13}$, $\theta$, $\phi$).

\subsection{Low Shear frequency}\label{sec:lowf}

For a low shear frequency ($f_{\gamma}^* \lesssim 0.1$),
the RBC can achieve the dynamics in the steady shear flow with $\dot\gamma \sim \dot\gamma_0$
for  a half period $1/2f_{\gamma}$. 
Therefore, at most of the parameter ranges, it approaches one limit-cycle oscillation from any initial position.
At the shear amplitude $\dot\gamma_0^* \gg \dot\gamma^*_{\rm {tt}}$ or $\dot\gamma_0^* < \dot\gamma^*_{\rm {tb}}$,
$\phi$ or $\theta$ rotates in the negative direction at $n < f_{\gamma} t < n+1/2$, respectively,
and rotates back to the original position at $n+1/2 < f_{\gamma} t < n+1$ (see  Figs. \ref{fig:cart} and \ref{fig:rq5}).
The shape parameter $\alpha_{13}$ and $\theta$ oscillate (swing) with the $\phi$ rotation frequency
at $\dot\gamma_0^* \gg \dot\gamma^*_{\rm {tt}}$.
This swinging amplitude decreases with increasing $\dot\gamma_0^*$.

At $\dot\gamma_0^* \sim \dot\gamma^*_{\rm {tt}}$,
both $\phi$ and $\theta$ can rotate, so
the RBC shows complicated behaviors,
which are sensitive to the parameters $\dot\gamma_0^*$ and $f_{\gamma}^*$.
It is found that intermittent and synchronized oscillations occur in the oscillatory flow (see Fig.~\ref{fig:itc}).
A typical intermittent oscillation is shown in the bottom-left panel of Fig.~\ref{fig:itc}.
The angles $\theta$ and $\phi$ occasionally rotate $\pm\pi$ with the shear frequency.
Synchronization of rotation with an $n$-fold shear-oscillation period is observed for a finite range of $f_{\gamma}^*$.
Thus, the Devil's staircase also appears in oscillatory shear flow.
The average number $\langle n_{\rm {rot}} \rangle$ of rotations increases 
as $\langle n_{\rm {rot}} \rangle \propto \sqrt{f_{\rm c}^*-f_{\gamma}^*}$ near the critical frequency $f_{\rm c}^*$.
This dependence indicates the type I intermittency \cite{berg84,skot07}.
These intermittency and synchronization are very similar to those in the steady flow~\cite{skot07,nogu09b}.
However, multiple attractors can coexist in the oscillator flow, unlike in steady flow.
When a trajectory is asymmetric, as shown in the bottom-right panel of Fig.~\ref{fig:itc},
one more trajectory exists.
The coexistence of four limit-cycle oscillations is also found at $\dot\gamma_0^*=0.002$ and $f_\gamma^*=0.014$ (data not shown).

\begin{figure}
\includegraphics{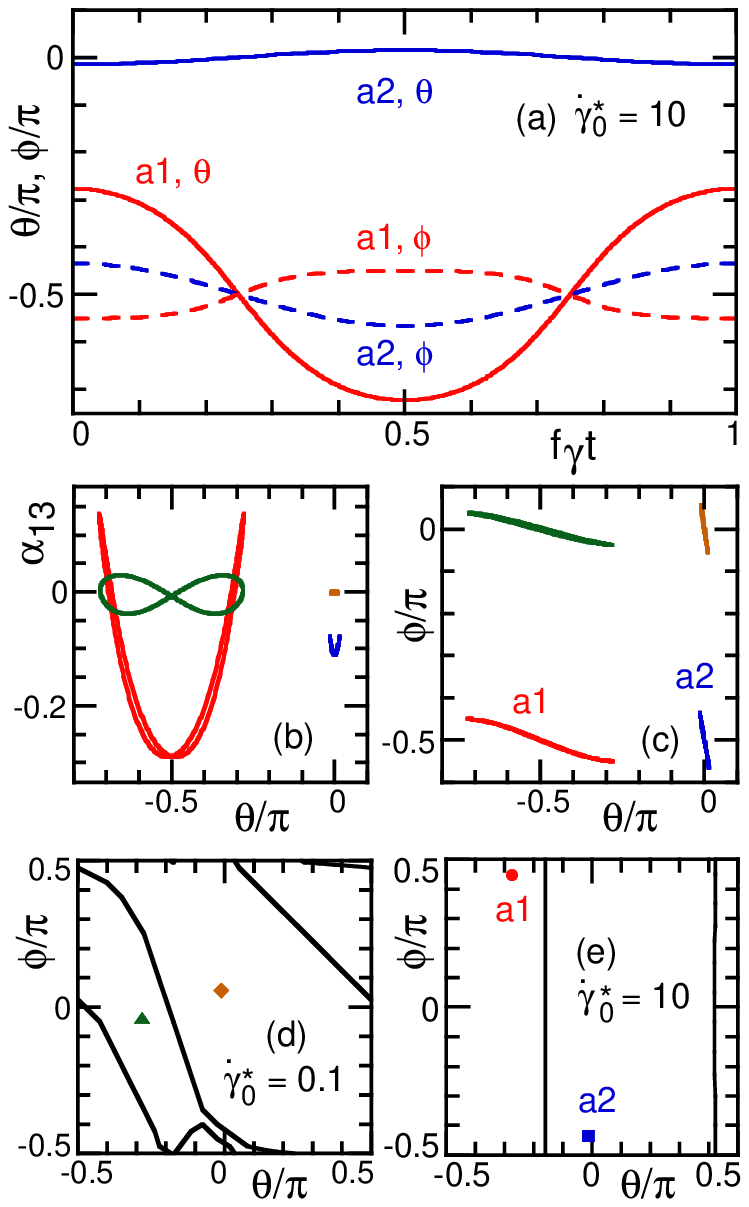}
\caption{ \label{fig:att}
(Color online)
RBC dynamics at high frequency $f_{\gamma}^*$ at $f_{\gamma}^*=0.2$.
(a) Time evolution of two limit-cycle oscillations at $\dot\gamma_0^*=10$.
 (denoted as a1 and a2).
(b), (c) Trajectories of the limit-cycles at $\dot\gamma_0^*=0.1$ and $10$.
(d), (e) Domains of the attractors (initial positions
 $(\alpha_{13}, \theta, \phi)=$
$(0, \theta_{\rm i}, \phi_{\rm i})$ at $t=0$) at (d) $\dot\gamma_0^*=0.1$ and (e) $10$.
Symbols represent the positions ($\theta, \phi$)
at $t=n/f_\gamma$ in the limit $n\to\infty$.
}
\end{figure}

\begin{figure}
\includegraphics{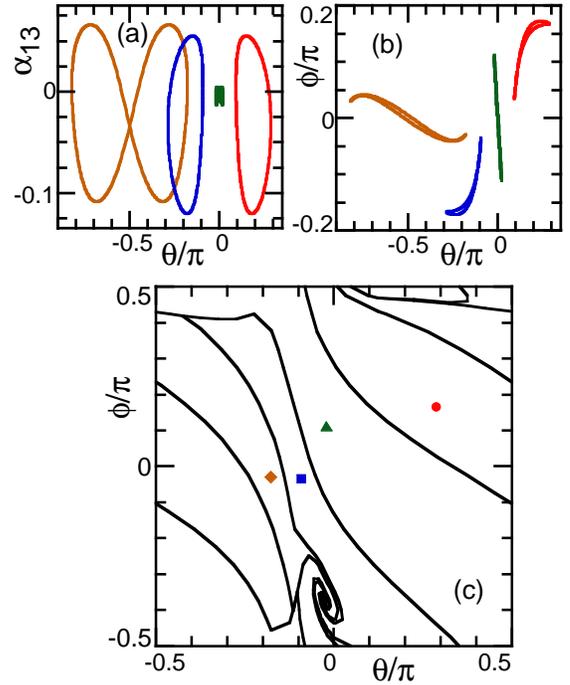}
\caption{ \label{fig:att4}
(Color online)
RBC dynamics at high frequency $f_{\gamma}^*$ at  $\dot\gamma_0^*=0.2$ and $f_{\gamma}^*=0.1$.
(a), (b) Trajectories of four limit-cycles.
(c) Domains of the attractors (initial positions
 $(\alpha_{13}, \theta, \phi)=$
$(0, \theta_{\rm i}, \phi_{\rm i})$ at $t=0$).
Symbols represent the positions ($\theta, \phi$)
at $t=n/f_\gamma$ in the limit $n\to\infty$.
}
\end{figure}

\begin{figure}
\includegraphics{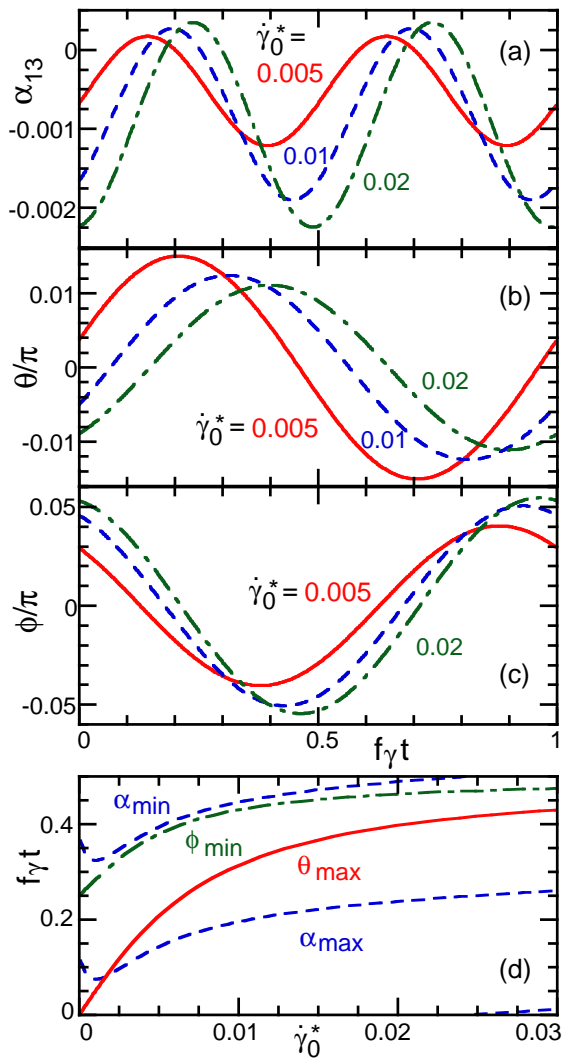}
\caption{ \label{fig:tb}
(Color online)
RBC dynamics at high frequency $f_{\gamma}^*=0.2$ with low $\dot\gamma_0^*$.
(a)--(c) Time evolution of limit-cycle oscillations around $\theta=0$.
Solid, dashed, and dashed-dotted lines represent $\dot\gamma_0^*=0.005$, $0.01$, 
and $0.02$, respectively.
(d) Shear amplitude $\dot\gamma_0^*$ dependence of
times $t$ at maximum or minimum of $\alpha_{13}$, $\theta$, and $\phi$.
The data for $0 \le t \le 0.5/f_\gamma$ is only shown
because $\alpha_{\rm D}(t+0.5/f_\gamma)=\alpha_{\rm D}(t)$, $\theta(t+0.5/f_\gamma)=-\theta(t)$,
  and $\phi(t+0.5/f_\gamma)=-\phi(t)$.
}
\end{figure}

\begin{figure}
\includegraphics{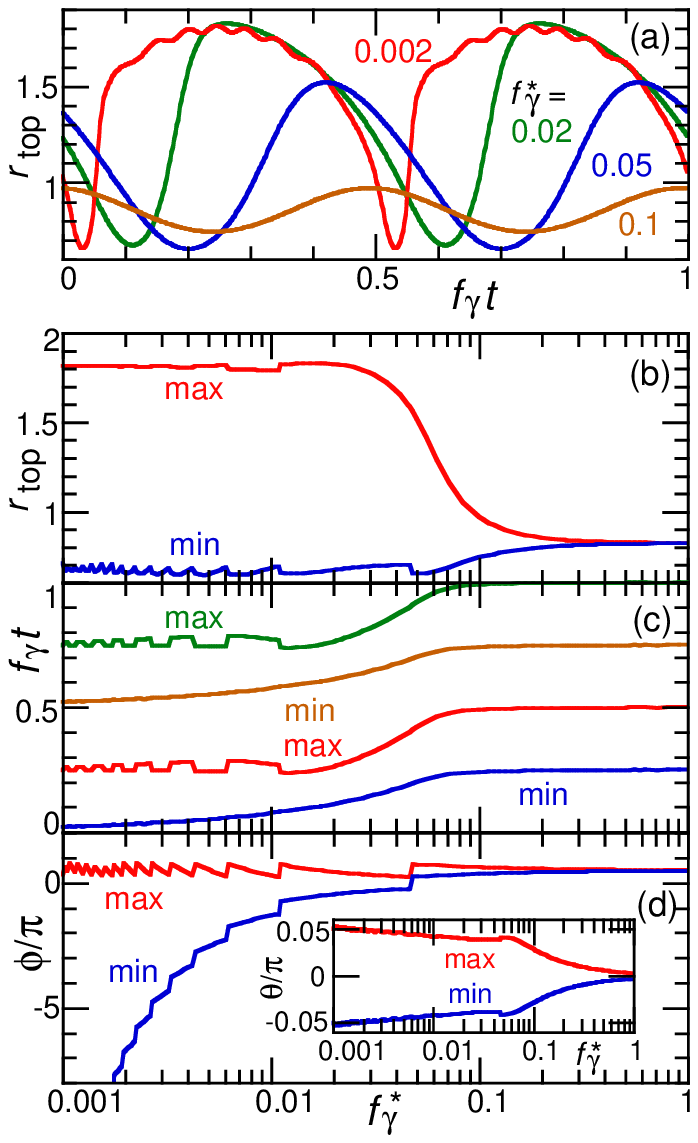}
\caption{ \label{fig:r4}
(Color online)
Dependence on shear frequency $f_{\gamma}^*$ at $\dot\gamma_0^*=10$
with a gradual increase in $f_{\gamma}^*$.
(a) Time evolution of the length ratio $r_{\rm {top}}=L_1 \cos(\theta)/L_3$
for various $f_{\gamma}^*$.
The frequency $f_{\gamma}^*$ dependence is shown for
(b) $r_{\rm {top}}$ and (c) $t$ at maxima and minima of the $r_{\rm {top}}$ curves in (a).
The maximum and minimum angles $\phi$ and $\theta$ are shown in (d) and the inset of (d), respectively.
}
\end{figure}

\subsection{High Shear frequency}\label{sec:highf}

For a high shear frequency ($f_{\gamma}^* \gtrsim 0.1$),
it is found that multiple ($2 - 4$) limit cycles coexist, as shown in Figs.~\ref{fig:atline}, \ref{fig:att}, and \ref{fig:att4}.
Since the shear frequency is higher than TT and TB frequencies,
$\phi$ or $\theta$ cannot fully rotate for $1/2f_{\gamma}$;
thus, multiple orbits are stabilized.
An approached limit cycle is chosen by initial angles ($\theta_i, \phi_i$)
but is almost independent of initial $\alpha_{13}$.
The shape parameter $\alpha_{13}$ relaxes much faster than $\theta$ and $\phi$.
As  $\dot\gamma_0^*$ increases,
it is less dependent on initial angle $\phi_i$,
and becomes almost independent of $\phi_i$ at $\dot\gamma_0^*=10$ [see Fig. \ref{fig:att}(e)],
since the energy barrier of the TT rotation becomes negligible at $\dot\gamma^* \gg \dot\gamma_{\rm {tt}}$.
At high or low $\dot\gamma_0^*$, two limit cycles can coexist [see Fig. \ref{fig:atline}(b)].
With increasing  $f_{\gamma}^*$, the domain for a new limit cycle appears at $\theta\simeq -0.2\pi$.
At high $\dot\gamma_0^*$,
in the limit cycle, which also exist in low $f_{\gamma}^*$, $\theta$ oscillates between $\pm \theta_{\rm {tt}}$,
where  $\theta_{\rm {tt}}$ is the angle in the steady flow with $\dot\gamma=\dot\gamma_0$.
In the other limit cycle, $\theta$ oscillates 
between $\theta_{\rm {tt}}$ and $\pi-\theta_{\rm {tt}}$ [see Fig. \ref{fig:att}(a)].

At low $\dot\gamma_0^*$ with $f_{\gamma}^* \gtrsim 0.1$,
two limit cycles coexist like at high $\dot\gamma_0^*$:
 $\theta$ oscillates between $\pm \theta_0$ or
  between $\theta_0$ and $\pi-\theta_0$ ($\theta_0 \simeq 0.1\pi$).
In the latter oscillation, $\theta$ decreases (increases) 
at $0<t<0.5/f_{\gamma}$ ($0.5/f_{\gamma}<t<1/f_{\gamma}$) like at high $\dot\gamma_0^*$ 
(see the solid line for a1 in Fig. \ref{fig:att}),
while the former oscillation around $\theta=0$ is different from that at high $\dot\gamma_0^*$.
At $\dot\gamma_0^* \to 0$, $\theta$ decreases at $0<t<0.5/f_{\gamma}$, and
has maximum and minimum at $t=0$ and $t=0.5/f_{\gamma}$, respectively, like for fluid vesicles \cite{nogu10}.
As $\dot\gamma_0^*$ increases, the times $t$ for the maximum and minimum of $\theta$
increase and approach $0.5$ and $1$, respectively (see Fig. \ref{fig:tb}).
Thus, $\theta$ rotates in the opposite direction to the shear despite of the TB phase.
This opposite rotation is induced by the temporal $\phi$ rotation in the shear direction.
The maxima and minima of $\alpha_{13}$ and $\phi$ also increase with increasing $\dot\gamma_0^*$.
The hight of the free-energy barrier for $\phi$ rotation
may be estimated from this $\dot\gamma_0^*$ dependence.

At the middle shear amplitudes $\dot\gamma_0^*=1 \sim 3$ with $f_{\gamma}^* \gtrsim 0.1$, 
the domains have a complicated shape.
Figure \ref{fig:att4} shows four limit cycle oscillations at  $(\dot\gamma_0^*,f_{\gamma}^*)=(0.2,0.1)$.
Two limit cycles show similar trajectories of those at $(\dot\gamma_0^*,f_{\gamma}^*)=(0.1,0.2)$;
compare Figs. \ref{fig:att}(b), (c) and Figs. \ref{fig:att4}(a), (b).
In addition to four stable fixed points,
 an unstable fixed point is seen at ($\theta$, $\phi$)=($-0.03\pi$, $-0.37\pi$)
in Fig. \ref{fig:att4}(c).
Around this unstable point, the angles move away from the unstable point 
with a spiral orbit and then approach one of the stable points.
As $f_{\gamma}^*$ increases, the domains of attractors merge or split [see Fig.~\ref{fig:atline}(d)].
These multiple cycles may not be desired for characterizing the mechanical properties in experiments.
However, one of the cycles is chosen when $f_{\gamma}$ is gradually increased
from the TT- or TB-based oscillation,
since there is only one cycle at low $f_{\gamma}$. 
Note that the approach to limit cycles is very slow at $f_{\gamma}^* \gtrsim 0.1$
and typically takes $f_{\gamma}t \sim 10^4$
($t= 10$ min to $1$ hour at $\dot\gamma_0^* \sim 1$).

\subsection{Dependence of Length Ratio}\label{sec:amp}

In the experiments in Ref.~\cite{wata06}, the length ratio $r_{\rm {top}}=L_x/L_3$ of RBCs from the top view
was measured at low frequency $f_{\gamma}^* \ll 1$, where $L_x$ is the length in the $x$ direction 
projected on the $xz$ plane. 
In TT-based oscillation, the ratio is approximated as
$r_{\rm {top}} = \cos(\theta)L_1/L_3=\cos(\theta)(1+\alpha_{13})/(1-\alpha_{13})$,
since RBCs are aligned in the $x$ direction with  $|\theta| \ll \pi$ [see the inset of Fig.~\ref{fig:r4}(d)].

At high $\dot\gamma_0^*$,
the $r_{\rm {top}}$ curves have the same shape at $n<f_{\gamma}t<n+1/2$ and $n+1/2<f_{\gamma}t<n+1$ 
[see Figs.~\ref{fig:r4}(a) and (c)].
At $f_{\gamma}^* \ll 1$,
RBCs have minimum or maximum deformation at $f_{\gamma}t = 0$ and $0.5$ or at $f_{\gamma}t \simeq 0.25$ and $0.75$, 
where the shear stress $\eta_0\partial v_x/\partial y=\eta_0 \dot\gamma$ is minimum or maximum, 
respectively.
As $f_{\gamma}^*$ increases, 
the oscillation amplitude decreases, and the times $t$ for the maximum and minimum deformations become delayed,
since the temporal change of the shear rate becomes faster than the shape relaxation (see Fig.~\ref{fig:r4}).
At $f_{\gamma}^* \simeq 1$,
the times $t$ for the maximum deformation approach $f_{\gamma}t = 0.5$ and $1$.
We cannot directly compare our results with the experiments \cite{wata06},
since the large shape deformations $r_{\rm {top}} \simeq 6$ ($\dot\gamma_0^*\sim 100$ and $f_{\gamma}^*=0.004$)
 in those experiments
 are beyond the range of the ellipsoidal-shape assumption $r_{\rm {top}} \le 5.3$ of the KS theory.
However, our $r_{\rm {top}}$ curve at $f_{\gamma}^*=0.004$ 
well reproduces those in Ref.~\cite{wata06}, except for the amplitude of $r_{\rm {top}}$.
Thus, we conclude that the shape oscillation observed in their experiments
is TT-based shape oscillation for a low frequency $f_{\gamma}^* \lesssim 0.1$.
Furthermore, our theoretical model predicts that TB-based or intermittent oscillations and multiple limit cycles
would occur for lower $\dot\gamma_0^*$ and higher $f_{\gamma}^*$, respectively.

\begin{figure}
\includegraphics{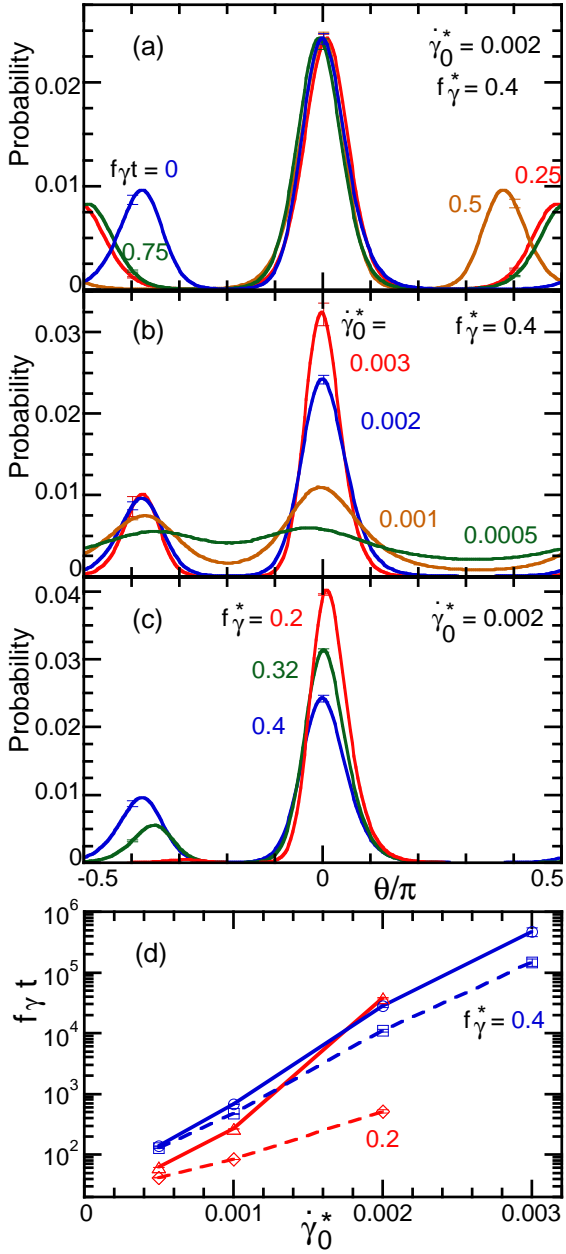}
\caption{ \label{fig:noise}
(Color online)
Thermal fluctuations effects at $\dot\gamma_0^* \ll 1$.
(a)--(c) Probability distributions of $\theta$. 
(a) Time evolution at $\dot\gamma_0^*=0.002$ and $f_{\gamma}/\dot\gamma_{\rm m}=0.4$:
$f_\gamma t=n$, $n+0.25$,  $n+0.5$, and $n+0.75$.
(b) Dependence on $\dot\gamma_0^*$ at $t=n/f_\gamma$ and $f_{\gamma}/\dot\gamma_{\rm m}=0.4$.
(c) Dependence on $f_{\gamma}^*$ at $t=n/f_\gamma$ and $\dot\gamma_0^*=0.002$.
(d) Mean lifetime of orbits aournd $\theta=0$ (solid line) and around $\theta=\pi/2$ 
(dashed line) at $f_{\gamma}^*=0.2$ ($\triangle$, $\diamond$) and $0.4$ ($\circ$, $\Box$).
The error bars are shown at (a)--(c) several and (d) all data points.
}
\end{figure}

\subsection{Thermal Fluctuations}\label{sec:noise}

In this subsection, we describe the effects of thermal fluctuations.
A dimensionless quantity, the rotational Peclet number,
$\chi =  \dot\gamma_0/D_\theta=\dot\gamma_0\zeta_{\theta}/k_{\rm B}T$
represents the shear amplitude relative to the thermal fluctuations.
In typical experimental conditions,
RBCs have very large  $\chi$ and
the thermal fluctuations are negligible;
 $\chi \simeq 2 \times 10^6 \dot\gamma_0^*$ at $\eta_0= 0.02$Pa$\cdot$s.
At  very low shear amplitudes $\dot\gamma_0^* \ll 1$,
however, the thermal fluctuations can induce large fluctuations of trajectories
and transitions between attractors.

Figure \ref{fig:noise} shows the dynamics with the thermal fluctuations at $\dot\gamma_0^* \ll \dot\gamma^*_{\rm {tt}}$  and high $f_{\gamma}^*$,
where two limit cycle orbits coexist in the absence of the thermal fluctuations.
Two peaks around $\theta=0$ and $\theta=\pi/2$ in Fig. \ref{fig:noise}(a) indicate that
these two orbits are still dominant with the thermal fluctuations.
The transitions between these orbits are obtained at very low $\dot\gamma_0^*$.
With increasing $\dot\gamma_0^*$,
the lifetime of each orbit exponentially increases and the transition probability exponentially decreases.
We calculate the lifetime as the duration from the time to enter the region of one orbit
to the time to enter the region of the other orbit,
where the regions of the orbits are considered as $-0.15<\theta/\pi<0.2$ and $0.4<\theta/\pi<0.75$.
As $f_{\gamma}^*$ decreases,
the domain of the orbit around $\theta=\pi/2$ is reduced [see Fig. \ref{fig:atline}(b)]. Also,
its lifetime decreases and the other orbit becomes dominant [see Figs. \ref{fig:noise}(c) and (d)].
Thus, attractors with small domain can be smeared out by the thermal fluctuations.
Since the lifetimes exhibit an exponential increase,
it would be very difficult 
to observe the transitions between orbits  in experiments.

\begin{figure}
\includegraphics{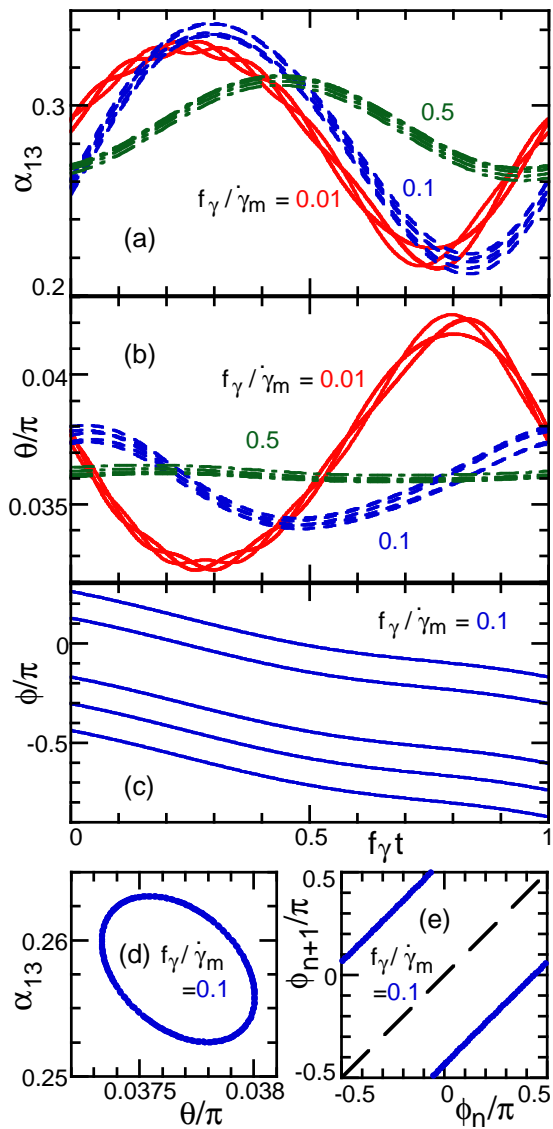}
\caption{ \label{fig:bias_rq}
(Color online)
Time evolution of (a) $\alpha_{13}$, (b) $\theta$, and (c) $\phi$ 
in the oscillatory flow with the mean shear rate $\dot\gamma_{\rm m}^*=10$ and
oscillatory amplitude $\dot\gamma_{\rm m}^*=5$.
(d) Stroboscopic map for $t=n/f_\gamma$ at $f_{\gamma}/\dot\gamma_{\rm m}=0.1$.
(e) Return map sampled stroboscopically for $t=n/f_\gamma$ at $f_{\gamma}/\dot\gamma_{\rm m}=0.1$.
Dashed line represent $\phi_{n+1}=\phi_{n}$.
The phase of the angle $\phi$ is not locked to the shear oscillation.
}
\end{figure}

\begin{figure}
\includegraphics{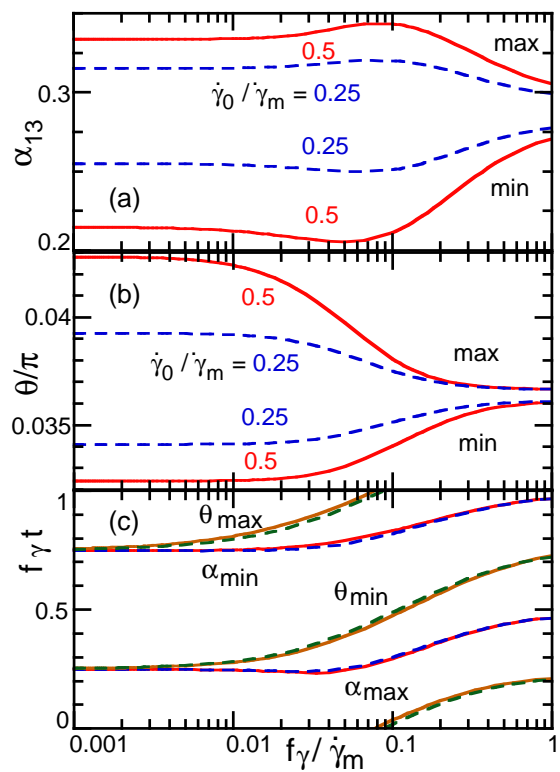}
\caption{ \label{fig:bias_amp}
(Color online)
Dependence on shear frequency $f_{\gamma}/\dot\gamma_{\rm m}$ at $\dot\gamma_{\rm m}^*=10$.
Maximum and minimum values of (a) $\alpha_{13}$ and (b) $\theta$. 
(c) Time $t$ at maximum and minimum of $\alpha_{13}$ and $\theta$.
Dashed and solid lines represent  $\dot\gamma_0/\dot\gamma_{\rm m}=0.25$ and $0.5$, respectively.
}
\end{figure}

\section{Oscillatory Flow with Finite Mean Shear Rate}\label{sec:bias}

When  oscillatory shear is applied with a finite mean shear rate $\dot\gamma_{\rm m}$ 
as $\dot\gamma(t)= \dot\gamma_{\rm m} + \dot\gamma_0 \sin(2\pi f_{\gamma} t)$,
a net rotation of $\theta$ or $\phi$ is obtained.
At high shear $\dot\gamma_{\rm m}^* -\dot\gamma_0^* \gg \dot\gamma_{\rm tt}^*$,
RBCs always show clockwise TT rotation accompanied by $\alpha_{13}$ and $\theta$ oscillations with shear frequency $f_{\gamma}$
(see Fig. \ref{fig:bias_rq}).
The coupling between the shear oscillation and $\phi$ rotation is very weak,
so that the synchronization can occur only in negligibly narrow ranges of $f_{\gamma}$.
Thus, $\phi$ typically rotates with its own frequency 
(the curves of $\phi$(t) and $\phi_{n+1}(\phi_{n})$ in Figs. \ref{fig:bias_rq}(c) and (e), respectively, 
are close to straight lines),
and  $\alpha_{13}$ and $\theta$ show swinging oscillations with the frequency of the $\phi$ rotation 
in addition to the oscillations with $f_{\gamma}$ [see Figs. \ref{fig:bias_rq}(a) and (b)]. 

As  $f_{\gamma}$ increases, the times $t$ for the maximum and minimum deformations become delayed
with respect to the times of the minimum and maximum shear stresses, while multiple limit cycles do not appear (see Fig. \ref{fig:bias_amp}).
A similar time delay of the shape deformation is experimentally observed in Ref. \cite{naka90}.
These time delays are determined by $f_{\gamma}/\dot\gamma_{\rm m}$ instead of $f_{\gamma}/\dot\gamma_0$.
When $\dot\gamma_0$ is varied, the oscillation amplitudes are changed, while the times for the minimum and maximum oscillations
are almost independent of $\dot\gamma_0$.
In the low frequency limit,
the maximum and minimum of $\alpha_{13}$ appear accompanied by minimum and maximum of $\theta$ at $t=0.25/f_\gamma$ and $t=0.75/f_\gamma$,
respectively.
With increasing  $f_{\gamma}$, the maximum and minimum of $\alpha_{13}$ approach $t=0.5/f_\gamma$ and $t=1/f_\gamma$, respectively,
where $\dot\gamma(t)=\dot\gamma_{\rm m}$.
A similar dependence is obtained for fluid vesicles \cite{nogu10}. 
The angle $\theta$ shows greater delays than $\alpha_{13}$,
since stable $\theta$ is varied not directly by $\dot\gamma^*$ but by the shape evolution.
Thus, the shape deformation is essential for the response to time-dependent flows.

\section{Summary and Discussion} \label{sec:sum}

We have investigated RBC dynamic modes in oscillatory shear flow for a wide range of the shear conditions.
For a low shear frequency ($f_{\gamma}^* \lesssim 0.1$) with zero mean shear rate,
RBCs exhibit TT- or TB-based oscillation at high or low shear amplitude $\dot\gamma_0^*$, respectively.
In the middle amplitude $\dot\gamma_0^*$,
intermittent or synchronized oscillations appear.
For  a high frequency ($f_{\gamma}^* \gtrsim 0.1$),
multiple limit-cycle oscillations appear.
Two limit cycles coexist for low and high $\dot\gamma_0^*$,
and two to four limit cycles coexist for middle $\dot\gamma_0^*$.
For a finite mean shear rate with small oscillation amplitudes,
 $\alpha_{13}$ and $\theta$ oscillate in addition to the swinging oscillation, and
there is only one attractor even at high $f_{\gamma}$.

In this paper, 
the symmetric axis of the RBC discoidal shape is assumed on the vorticity ($xy$) plane.
This assumption is valid in most of the parameter range (including the parameters in the present study).
However, a few studies were reported on
a spinning motion (a principal axis rotates out of the  vorticity plane) in steady shear flow.
Lebedev {\it et al.} predicted that fluid vesicles exhibit the spinning motion
at very large $\dot\gamma^*$ and large $\eta_{\rm {in}}^*$ using the perturbation theory for quasi-spherical vesicles \cite{lebe08}.
Bitbol observed that the symmetric axis of the RBC discoidal shape is oriented in the vorticity ($z$) direction at low $\dot\gamma^*$
and large internal viscosity $\eta_{\rm {in}}^* = 1 \sim 10$ \cite{bitb86}.
In the oscillatory flow, RBCs may exhibit spinning dynamics at large $\eta_{\rm {in}}^*$
but it is beyond the scope of our present study.
 
For fluid vesicles in the oscillatory shear flow with zero mean share rate,
the bifurcation frequency $f_\gamma^*$ to start coexistence of two limit cycles
decreases with increasing $\eta_{\rm {in}}^*$  in the TT phase,
since the average angular velocity $\dot\theta$ to relax to stable angle becomes slower \cite{nogu10}.
A similar dependence on $\eta_{\rm {in}}^*$ is expected in RBC dynamics.
Since the membrane and internal fluid become more viscous on aging of RBCs\cite{tran84,nash83},
the bifurcation frequency can be shifted on aging.

Recently, the relation of the dynamic modes of RBCs or vesicles to the viscosity of a dilute suspension
was studied \cite{vitk08,kant08}. The dependence of storage and loss moduli of the dilute suspension \cite{lars99}
on the dynamic modes in the oscillatory shear flow is also an interesting problem for further studies.
In high frequencies $_{\gamma}^*$, the collisions between RBCs may induce a transition between coexisted limit-cycle orbits
as the thermal fluctuations at very low $\dot\gamma_0^*$.

Watanabe {\it et al.} \cite{wata06,wata07a} proposed that the response curve of $r_{\rm {top}}$ at low $f_{\gamma}^*$ is a good
quantity for evaluating RBC deformability.
Experimental measurement of the dynamic response for a wide range of $\dot\gamma_0^*$ and $f_{\gamma}^*$
would be a significant help in establishing a quantitative understanding of the mechanical properties of RBCs,
in particular the viscoelasticity of RBC membrane.
Changes of RBC deformability in different diseases may be able to separated by their dynamic response.
We applied the model to RBCs, but
the resulting dynamics would also occur for other elastic capsules by the modification of the free-energy potential $F$.
The oscillatory shear flow is a very useful setup
for measuring the viscoelasticity of RBCs and other soft deformable objects such as synthetic capsules and lipid vesicles.

\begin{acknowledgments}
We thank 
N. Watanabe (Tokyo Med. Dental Univ.) for the helpful discussion.
This study is partially supported by a Grant-in-Aid for 
Scientific Research on Priority Area ``Soft Matter Physics'' from
the Ministry of education, Culture, Sports, Science, and Technology of Japan.
\end{acknowledgments}


\end{document}